\documentstyle[twoside,fleqn,espcrc2,epsf]{article}
\epsfverbosetrue
\newcommand{\err}[2]{{\small {$\;\begin{array}{@{}l@{}}
			  +\makebox[0.9em][r]{#1} \\[-0.4em]
			  -\makebox[0.9em][r]{#2}
			\end{array}$}}}
\newcommand{\ewxy}[2]{\setlength{\epsfxsize}{#2}\epsfbox[10 60 640 570]{#1}}
\def\slash#1{\setbox0=\hbox{$#1$}               % set a box for #1
        \dimen0=\wd0                            % and get its size
        \setbox1=\hbox{/} \dimen1=\wd1          % get size of /
        \ifdim\dimen0>\dimen1                   % #1 is bigger
        \rlap{\hbox to \dimen0{\hfil/\hfil}}    % so center / in box
        #1                                      % and print #1
        \else                                   % / is bigger
        \rlap{\hbox to \dimen1{\hfil$#1$\hfil}} % so center #1
        /                                       % and print /
        \fi}                                    %
\title{ Gauge Invariant Smearing and the Extraction of Excited State
Masses Using Wilson
Fermions at $\beta=6.2$}

\author{ {\large \it UKQCD
Collaboration} - presented by Sara Collins\\
Department of Physics, The University of Edinburgh, Edinburgh EH9~3JZ,
Scotland}
\begin{document}

\begin{abstract}

We present an investigation of gauge invariant smearing for Wilson
fermions on a $24^3 \times 48$ lattice at $\beta = 6.2$. We
demonstrate a smearing algorithm that allows a substantial improvement
in the determination of the baryon spectrum obtained using propagators
smeared at both source and sink, at only a small computational cost.
We investigate the matrix of correlators constructed from local and
smeared operators, and are able to expose excited states of both the
mesons and baryons.

\end{abstract}
\maketitle
\section{Introduction}
In this talk we present the results of an investigation into gauge
invariant smearing on a $24^3\times48$ lattice at $\beta=6.2$ with a
hopping parameter $\kappa=0.152$. This choice of $\kappa $ corresponds to a
pseudoscalar
meson mass of approximately $600$ MeV if we take the scale from the
string tension ($a^{-1}\sim2.73(5)$ GeV~\cite{hadrons_letter}). We
begin by comparing effective masses obtained from local correlators
with those from Wuppertal or ``scalar wave function'' smeared
correlators~\cite{guesken90}. We introduce a variant of the iterative
Wuppertal scheme which enables us to achieve the results of scalar wave
function smearing but at a much lower cost in computer time. Using this
scheme, we perform detailed measurements of the hadron spectrum
using local and smeared sources and sinks. Finally,
we investigate the $2 \times 2$ matrices of correlators formed from
propagators computed with these four source-sink combinations, and by
diagonalising these matrices are able to isolate the first excited states.
%\vspace{-0.3cm}
\section{Gauge Invariant Smearing}
\subsection{Wuppertal Smearing}
\begin{figure}
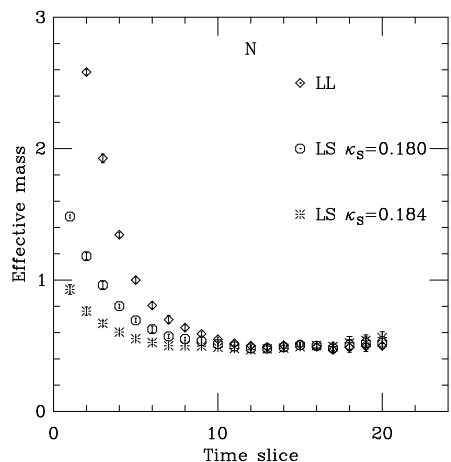

\centerline{\ewxy{wuppertal.ps}{70mm}}
\caption{A comparison between the effective mass of the nucleon obtained
using LL and LS propagators.
\label{fig:eff_mass_test}}
\end{figure}

Wuppertal smearing uses a scalar propagator to smear a delta function source at
the
origin~\cite{guesken90}. The smeared source is
given by the solution of the three dimensional, gauge invariant, Klein-Gordon
equation.
For our initial investigation of the dependence of the effective masses
on the smearing radius, quark propagators were computed on an ensemble
of 7 configurations using Wuppertal smeared sources at $\kappa_S =
0.180$ and $0.184$, corresponding to $r \simeq 2$ and $r \simeq 4$
respectively~\cite{smearing_paper}. The effective mass of the nucleon obtained
using LL and
LS propagators on the same ensemble of configurations is shown in
figure~\ref{fig:eff_mass_test} (we adopt the convention of LS
representing a propagator smeared at the source and local at the sink).
The lightest state of the nucleon is isolated nearer the origin with increasing
smearing radius.
%
%  here we insert the figures of the effective masses

The Wuppertal smearing algorithm creates significant computational overheads
when
used to smear the propagator at the sink; the
smearing algorithm must be implemented on every time slice, not just
the source time slice, and for every spin component. Indeed, for
$\kappa_S = 0.184$, the computational effort required to smear at the
sink is comparable with that required to compute the propagator, and
larger values of the smearing radius are prohibitively expensive.
\subsection{Jacobi Smearing}
  An alternative smeared source, $J(x)$, can be obtained by solving the three
dimensional Klein-Gordon equation as a power series in $\kappa_{S}$,
stopping at some finite power $N$.
\begin{eqnarray}
 J(x) & = & \sum_{n=0}^{N} \kappa_S^{^n}\Delta^{^n}\delta_{x,0}
\end{eqnarray}
where
\begin{eqnarray}
 \Delta_{x,x'} & = & \sum_{\mu=1}^{3} U_{\mu}(x)
\delta_{x',x+\mu}+U^{\dagger}_{\mu}(x-\mu)\delta_{x',x-\mu} \nonumber
\end{eqnarray}
This can be achieved easily using
Jacobi iteration. When $\kappa_{S}$ is smaller than some critical
value, the power series converges and one recovers the scalar wave
function. For sufficiently large $\kappa_{S}$, the series diverges but
provides an acceptable smeared source for suitable choices of $N$. A similar
iterative scheme has been used by the Wuppertal
group~\cite{guesken90,alexandrou91}.

 We found that for a radius of approximately 4, smearing  with the
Jacobi iteration was a factor of ten faster. The remainder of this paper will,
therefore, concentrate on results obtained using $\kappa_S = 0.250, N =
50$, corresponding to $r \simeq 4$, on the same 18 configurations which
were used in~\cite{hadrons_letter}.
%The nucleon effective mass
%obtained using the LL, LS, SL and SS propagators are shown in
%figure \ref{eff_mass_all}.
%% %  here we insert the figures of the effective masses %
%\begin{figure}
%\centerline{\ewxy{jacobi.ps}{70mm}}
%\caption{Smearing Function \label{fig:spectral-rep}}
%\end{figure}
\begin{table}[t]
% space before first and after last column: 1.5pc
%space between columns: 1.5pc (twice the above)
\setlength{\tabcolsep}{0.25pc}
%-----------------------------------------------------
% adapted from TeX book, p. 241
\newlength{\digitwidth}
\settowidth{\digitwidth}{\rm0}
\catcode`?=\active \def?{\kern\digitwidth}
%-----------------------------------------------------
\caption{The estimates of baryon and meson masses}
\label{tab:hadron_masses}
\begin{tabular}{lcccc}
                 & \makebox[0pt][c]{LL(12-16)}
                 & \makebox[0pt][c]{LS(11-16)}
                 & \makebox[0pt][c]{SL(11-16)}
                 & \makebox[0pt][c]{SS(7-12)}         \\
\hline
$m_{\pi}$    & $0.223\mbox{\err{5}{6}}$ & $0.219\mbox{\err{9}{4}}$   &
$0.220\mbox{\err{12}{8}}$  & $0.220\mbox{\err{7}{7}}$ \\
$m_{\rho}$ & $0.341\mbox{\err{10}{7}}$ & $0.327\mbox{\err{8}{7}}$ &
$0.356\mbox{\err{21}{13}} $  & $0.327\mbox{\err{9}{8}}$ \\
$m_{N}$   & $  0.510\mbox{\err{23}{10}}$ & $0.495\mbox{\err{12}{12}}$  &
$0.507\mbox{\err{31}{24}}$  & $0.495\mbox{\err{9}{10}}$ \\
$m_{\Delta}$  & $0.595\mbox{\err{15}{16}}$ & $0.567\mbox{\err{12}{17}}$  & $
0.593\mbox{\err{32}{16}}$    & $ 0.559\mbox{\err{9}{10}}$ \\
\hline
\end{tabular}
\end{table}
%\vspace{-0.3cm}
\section{Hadron Masses}
We present in table 1 the masses of the pseudoscalar, vector, nucleon and
$\Delta$ using the LL, LS, SL and SS propagators, together with the
corresponding time ranges used for the fits. All fits are performed
using the full covariance matrix, with the errors extracted using a
bootstrap analysis \cite{hadrons_paper}. The
improvement in the determination of the baryon masses through the use
of smeared sources and sinks is substantial. We expect that the study
of a larger sample of configurations would lead to a still more marked
improvement through the ability to perform a fit to the full
covariance matrix for the SS correlators over a larger fitting range
than that quoted in table 1. The lack of
improvement in the determination of the meson masses suggests that
mesons are already well exposed using local sources and sinks.

A recent analysis~\cite{daniel92} of the hadron spectrum obtained
using both wall sources and Wuppertal sources suggests that there is a
systematic difference in the baryon masses in the two cases, and that
this difference is particularly noticeable in the determination of the
$\Delta$ mass.
To test for such a discrepancy in our data, we perform
a bootstrap analysis of the $\Delta$ mass differences for the various
combinations of sources and sinks, using the time ranges of
table 1. For the LS and SL correlators, we find
\begin{equation}
m_\Delta^{LS} - m_\Delta^{SL} = 0.026\mbox{\err{31}{20}}\hspace{0.1cm}.
\end{equation}
This discrepancy is a $1 \sigma $ effect. However, since
the expectation values of the LS and SL correlators should be
identical, we attribute the discrepancy to limited statistics.
\section{Matrix Correlators}
By studying the matrix of correlators formed using the local and smeared
operators, ${\cal O}_{L}$ and ${\cal O}_{S}$ respectively, we can attempt
to extract masses for the first excited state.
We define the $2 \times 2$ matrix, $C(t)$, of timeslice correlators by
\[
	C(t) = \left( \begin{array}{cc}
		c_{LL}(t) & \omega c_{LS}(t) \\
		\omega c_{SL}(t) & \omega^2 c_{SS}(t)
		\end{array} \right)
\]
with
\begin{equation}
	c_{ij}(t) = \sum_{\vec{x}} \langle 0 |
		{\cal O}_i(\vec{x},t) {\cal O}_j^{\dagger}(\vec{0},0)
		| 0 \rangle .
		\label{C_definition}
\end{equation}
The factor $\omega$ arises from the arbitrary normalisation of the
smearing function.
Inserting a complete set of states in equation~(\ref{C_definition}), we
obtain
\begin{equation}
c_{ij}(t) = \sum_{n=0}^{\infty} \langle 0 |{\cal O}_i(\vec{0},0) | n
\rangle \langle n | {\cal O}_j^{\dagger}(\vec{0},0) | 0 \rangle \frac{e^{-E_n
t}}{2 E_n}.
\end{equation}
We will now discuss two methods for studying the matrix $C(t)$~\cite{luscher}.

\subsection*{Diagonalisation of the Transfer Matrix}
Consider the eigenvalue equation
\begin{equation}
C(t) u = \lambda(t,t_0) C(t_0) u.
\label{eq:luscher}
\end{equation}
for fixed $t_0$. If the system
comprises only two independent states, then the eigenvalues of
eq.~\ref{eq:luscher} are
\begin{eqnarray}
\lambda_{+}(t,t_0) & = & e^{-(t - t_0) E_0} \nonumber \\
\lambda_{-}(t,t_0) & = & e^{- (t - t_0)E_1}.
\label{eq:luscher_eigens}
\end{eqnarray}
The two states are
separated exactly, and the coefficients in eq.~\ref{eq:luscher_eigens}
grow exponentially with $t_0$. In general, where there are more than
two states, for sufficiently large $t$ we expect two states to be
dominant. The coefficients of the contributions of the higher states
to eq.~\ref{eq:luscher_eigens}  do not exhibit exponential growth with
$t_0$~\cite{luscher}. Hence ideally we wish to study $\lambda_{-}(t,t_0)$
for $t_0$ as large as possible. However, the increase in the noise in
the data far from the source generally requires that we choose $t_0$
close to the origin.
\subsection*{Diagonalisation of $C(t)$}
We can compute the eigenvalues, $\chi_{+}(t,\omega)$ and $\chi_{-}(t,\omega)$,
of $C(t)$
directly, and at large times, $t$, obtain
\begin{eqnarray}
\chi_{+}(t,\omega) & = & f_{+}(\omega) e^{-E_0 t}
	\{ 1 +
  g_{+}(\omega) O(e^{- \Delta E_1 t}) \}
	\nonumber \\
\chi_{-}(t,\omega) & = & f_{-}(\omega)
	e^{- E_1 t} \{ 1 + g_{-}(\omega)O(e^{ - \Delta \tilde{E} t}) \}, \nonumber \\
& &
\label{eq:sara_eigens}
\end{eqnarray}
where
\begin{eqnarray}
\Delta E_1 & = & E_1 - E_0 \nonumber \\
\Delta \tilde{E} & = & \mbox{min}[E_1 - E_0,
	E_2 - E_1].
\end{eqnarray}

As in the previous method, the eigenvalues $\chi_{+}(t,\omega)$ and
$\chi_{-}(t,\omega)$ are dominated by the ground state and first
excited state respectively. However, even when there are only two
states, corrections to the leading behaviour of the eigenvalues remain.
The coefficients of these corrections depend on the arbitrary parameter
$\omega$ and, since we diagonalise $C(t)$ at each timeslice, these
coefficients are not positive definite. We will exploit the dependence
on $\omega$ to seek cancellations between contributions from higher
states, and hence extend the plateau region in the
effective mass closer to the source.

Such an approach has clear dangers. In particular, the effective
masses can approach their asymptotic values either from above or from
below as we vary $\omega$. One restriction does temper the
uncontrolled nature of this method. We observe that for sufficiently
large times the excited state contributions to
eq.~\ref{eq:sara_eigens} will be negligible, and thus the effective
masses derived from $\chi_{+}(t,\omega)$ and $\chi_{-}(t,\omega)$ must be
insensitive to $\omega$.

\subsection{Results}
\begin{figure}
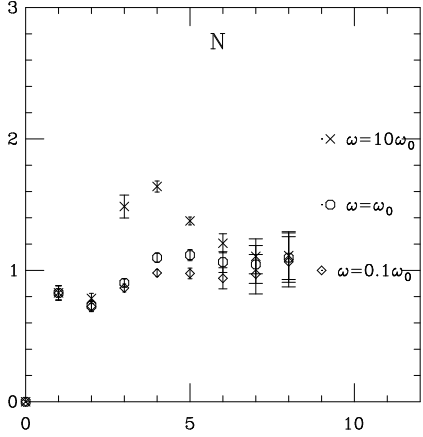

\centerline{\ewxy{Nuc_omega_paper.ps}{70mm}}
\caption{The variation with $\omega$ of the effective mass derived from
$\chi_{-}(t,\omega)$
for the first excited state of the nucleon. All values of $\omega$ are
quoted relative to $\omega_0$, where $\omega_0$ is defined as
$\protect\sqrt{\frac{c_{LL}}{c_{SS}}}$ at $t=12$.
\label{fig:lambda_minus_plots}}
\end{figure}

In figure \ref{fig:lambda_minus_plots}, we show the effective mass
derived from  $\chi_{-}(t,\omega)$ for the first excited state of the nucleon
at several values of
$\omega$. We observe there is a region in $t$ for which the effective
masses coincide for all $\omega$. As noted above this can be taken
as a signal that the first excited state has been isolated. The plateau
region in the effective mass appears to extend closer to the source
as $\omega$ decreases. We attribute this to cancellations
between contributions of the higher excited states to $\chi_{-}(t,\omega)$.

\begin{figure}
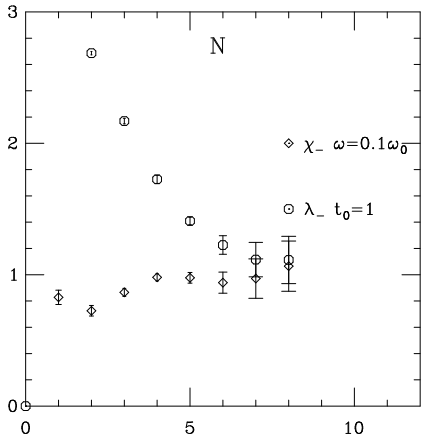

\centerline{\ewxy{Nuc_comp_paper.ps}{70mm}}
\caption{The effective mass of the first excited state of the nucleon derived
from
$\lambda_{-}(t,t_0)$, and $\chi_{-}(t,\omega)$.
\label{fig:plot_comparison}}
\end{figure}

In figure~\ref{fig:plot_comparison} we show the effective mass
of the first excited state of the nucleon derived from
$\lambda_{-}(t,t_0)$ with $t_0=1$, and
derived from $\chi_{-}(t,\omega)$ at the
optimal value of $\omega$. The two methods yield consistent plateau,
and using Method~2 we obtain
\begin{eqnarray}
\tilde{m}_{N} (4 - 8) & = & 0.98\mbox{\err{2}{4}}\hspace{.1cm},
\label{eq:excited_mass_fits}
\end{eqnarray}
where the fitting range is shown in brackets.

Comparing the computed ratio of the mass of the first excited state and
that of the ground state for the nucleon with its physical value for light
quarks,
we find it is appreciably larger. We find $\tilde{m}_N
/ m_N \simeq 2.0$, whereas $m_{N(1440)} / m_N = 1.5$. If this is
disappointing, it should be noted that a mass of $O(a^{-1})$ is likely
to be subject to considerable uncertainties arising from the non-zero
lattice spacing. However, an unexpectedly large value for the mass of the
nucleon excited state has been observed in other
simulations~\cite{marinari92}.
%\vspace{-0.3cm}
\section{Acknowledgements}
This research is supported by the UK Science and Engineering Research
Council under grants GR/G~32779 and GR/H~49191, by the University of
Edinburgh and by Meiko Limited.	We are grateful to Mike Brown of
Edinburgh University Computing Service, and to Arthur Trew and Paul
Adams of EPCC, for provision and maintenance of service on the Meiko
i860 Computing Surface and the Thinking Machines CM-200.
%\vspace{-0.3cm}

\end{document}